\documentclass[useAMS,usenatbib,usegraphicx]{mn2e}
\newcommand{\der}{{\rm d}}
\newcommand{\eq}{_{\rm e}}
\newcommand{\m}{_{\rm m}}
\newcommand{\s}{_{\rm s}}
\newcommand{\f}{_{\rm f}}
\newcommand{\so}{_{\rm s,0}}

\newcommand{\rs}{r_{\rm s}}
\newcommand{\rss}{\bmath{r_{\rm s}}}
\newcommand{\mss}{\bmath{M_{\rm s}}}
\newcommand{\rso}{r_{\rm s,0}}
\newcommand{\rsmin}{r_{\rm s,min}}
\newcommand{\rsmax}{r_{\rm s,max}}
\newcommand{\rsm}{r_{\rm s,m}}
\newcommand{\ti}{t_{\rm i}}
\newcommand{\tp}{_{\rm tp}}
\newcommand{\fa}{_{\rm a}}

\newcommand{\ra}{r_{\rm a}}

\newcommand{\PS}{_{\rm PS}}

\newcommand{\roc}{\rho_{\rm c}}

\newcommand{\delm}{\Delta_{\rm m}}

\newcommand{\delv}{\Delta_{\rm v}}

\newcommand{\nbody}{{$N$}-body }

\newcommand{\vc}{v_{\rm c}}

\begin{document}


\title[Halo Scale Radii and Mass Aggregation Histories]
{Scale Radii and Aggregation Histories of Dark Haloes}


\author[Salvador-Sol\'e, Manrique \& Solanes]
{Eduard Salvador-Sol\'e\thanks{E-mail: e.salvador@ub.edu},
Alberto Manrique and Jos\'e Mar\'\i a Solanes \\
Departament d'Astronomia i Meteorologia and Centre Especial de Recerca en
Astrof\'\i sica, F\'\i sica de Part\'\i cules i Cosmologia
\thanks{Associated with the Instituto de 
Ciencias del Espacio, Consejo Superior de Investigaciones Cient\'\i ficas} \\
Universitat de Barcelona, Mart{\'\i} i Franqu\`es 1, E-08028
Barcelona, Spain} 
%\email{e.salvador@ub.edu, a.manrique@ub.edu,jm.solanes@ub.edu}

%% are identified by the \altaffilmark after each name. Specify alternate


\maketitle
\begin{abstract}
Relaxed dark-matter haloes are found to exhibit the same universal
density profiles regardless of whether they form in hierarchical
cosmologies or via spherical collapse. Likewise, the shape parameters
of haloes formed hierarchically do not seem to depend on the epoch in
which the last major merger took place. Both findings suggest that the
density profile of haloes does not depend on their aggregation
history. Yet, this possibility is apparently at odds with some
correlations involving the scale radius $\rs$ found in numerical
simulations. Here we prove that the scale radius of relaxed,
non-rotating, spherically symmetric haloes endowed with the universal
density profile is determined \emph{exclusively} by the current values
of four independent, though correlated, quantities: mass, energy and
their respective instantaneous accretion rates. Under this premise and
taking into account the inside-out growth of haloes during the
accretion phase between major mergers, we build a simple physical
model for the evolution of $\rs$ along the main branch of halo merger
trees that reproduces all the empirical trends shown by this parameter
in \nbody simulations. This confirms the conclusion that the empirical
correlations involving $\rs$ do not actually imply the dependence of
this parameter on the halo aggregation history. The present results
give strong support to the explanation put forward in a recent paper
by Manrique et al.~(2003) for the origin of the halo universal density
profile.
\end{abstract}


\begin{keywords}
cosmology: theory -- dark matter -- galaxies: haloes
\end{keywords}


\section{INTRODUCTION}\label{intro}

High-resolution \nbody simulations of hierarchical cosmologies with
standard cold-dark-matter show that the spherically averaged density
profile of relaxed dark haloes in the present epoch is universal. It
is always well fitted by the analytical function 
\begin{equation}
\rho(r)=\frac{\roc\rs^3}{r^\beta(\rs+r)^{3-\beta}}
\end{equation}
where $r$ is the radius, $\beta$ the central asymptotic logarithmic
slope, with fixed value equal to 1 according to NFW (see also, e.g.,
\citealt{HJS99,ji00,Bull01,P03}) or closer to 1.5 according to other
authors (e.g., \citealt{Moore98,Gea00,FM01,K01}), and $\roc$ and $\rs$
the scaling density and radius. Thus, halos with a given mass have
density profiles characterised by one single parameter. In fact, the
scale radius $\rs$ or the concentration $c$, defined as $R/\rs$ in
terms of the virial radius $R$, are found to tightly correlate with
halo mass in all the cosmologies studied by NFW, a relationship that
is believed to reflect the fact that less massive haloes typically
form earlier, when the mean cosmic density is higher (NFW;
Salvador-Sol\'e, Solanes \& Manrique 1998). \citet{Bull01}, Wechsler
et al.~(2002, hereafter WBPKD), and Zhao et al.~(2003a, hereafter
ZMJB) have both extended these results to other redshifts in the flat
$\Lambda$CDM cosmology (see also \citealt{ENS01}; Zhao et al.~2003b;
\citealt{Tat04}) and provided toy-models for the evolution of $\rs$.

Within the hierarchical paradigm of structure formation, mergers
appear to contribute decisively to the evolution of haloes. For this
reason, it seems natural to try to explain the origin of the universal
halo density profile focusing on the role of mergers (e.g.,
Salvador-Sol\'e et al.~1998; \citealt{sw98}; \citealt{rgs98};
\citealt{SCO00}; \citealt{DDH03}). While a number of investigations
\citep{arfh,ns99,pgrs,kull00,wbd04} have demonstrated that NFW-like
density profiles can be obtained by simple spherical collapse, it is
not clear why this functionality should be preserved after major
mergers since these events produce an important loss of memory through
the rearrangement and violent relaxation of the system. Yet, as shown
by numerical experiments, haloes end up with very similar density
profiles regardless of whether or not they have undergone major
mergers \citep{HJS99} and, provided they have, regardless of the epoch
when the last of these events took place (\citealt{Moore99};
WBPKD). The situation is even more confusing when one takes into
account that the above mentioned mass-concentration relation and other
correlations involving $\rs$ (or $c$) recently reported by WBPKD and
ZMJB (see \S~\ref{sim}) suggest, on the contrary, that the structure
of haloes depends on their individual aggregation\footnote{Throughout
the present paper, we use the word ``aggregation'' to refer to halo
growth through mergers of any strength and reserve the word
``accretion'' for growth due to minor mergers only.} history.

In a recent paper, Manrique et al.~(2003, hereafter MRSSS) have shown
that the sole hypothesis that haloes grow inside-out with the typical
accretion rate of the cosmology under consideration automatically
leads to a density profile of the NFW-like form as well as to the
mass-concentration relation found at $z=0$ by NFW in numerical
simulations of that cosmology. Moreover, the mass-concentration
relation predicted for the $\Lambda$CDM universe at different
redshifts is also in good agreement with the empirical correlations
obtained by Bullock et al.~(2001) (MRSSS; \citealt{Hio03}). MRSSS
argued that the only reasonable explanation for the good predictions
of this pure inside-out accretion scheme is that the typical density
profile of relaxed haloes is essentially determined by their current
mass and the boundary conditions set by the matter that is currently
being accreted. In this manner, \emph{relaxed haloes emerging from
major mergers would have radial profiles indistinguishable from those
of haloes with identical global properties and boundary conditions but
having endured a more gentle growth}. This would also explain the
fact pointed out by \citet{Gea04} that some aspects of massive
elliptical galaxies imply that haloes evolve by preserving their inner
structure as in the pure inside-out accretion scheme \citep{LP03}
while they actually experience important rearrangements in major
mergers. However, this possibility is apparently in contradiction
with the correlations involving $\rs$ (or $c$) found in numerical
simulations.

In the present paper we examine this issue in detail. In
\S~\ref{sradius}, we show that, under some common approximations,
$\rs$ depends exclusively on the current values of the halo total
mass, energy and the respective instantaneous accretion rates, so
there is no room for a dependence on the halo aggregation
history. From this result, we develop, in \S~\ref{model}, a simple
physical model for the evolution of the scale radius along the main
branch of halo merger trees that allows us to show, in \S~\ref{sim},
that there is actually no contradiction between the independence of
$\rs$ from the halo aggregation history and the correlations found in
numerical simulations suggesting the opposite conclusion. Our results
are summarised and discussed in \S\ \ref{dis}. Unless otherwise
stated, the cosmology used throughout this paper is the same as in
WBPKD and ZMJB studies, that is, a $\Lambda$CDM model with
$(\Omega\m,\Omega_\Lambda,h,\sigma_8)=(0.3,0.7,0.7,1)$.

\section{GENERAL DEPENDENCE OF THE SCALE RADIUS}\label{sradius}

We will assume hereafter that relaxed haloes are non-rotating and
spherically symmetric (see \S\ \ref{dis} for a discussion on the
implications of these first-order approximations). Taking the
potential origin such that $\Phi(R)=-GM/R$, the virial relation adopts
the simple scalar form
\begin{equation}
\int_0^R \der M(r)\,\frac{M(r)}{r}= \frac{-2E+4\pi\,R^3P}{G}\,,
\label{vir}
\end{equation}
where $G$ is the gravitational constant, $E$ the total energy of the
halo, $P$ the pressure at the virial radius $R$ and $M(r)$ the mass
within $r$,
\begin{equation}
M(r)=\int_0^r \der \tilde r\, 4\pi \tilde{r}^2\rho(\tilde r)\,.
\label{mass}
\end{equation}

Since the virial radius of a halo at a given cosmic time $t$ is a
function of its total mass through the relation
\begin{equation}
R=\left[ {3\,M \over 4 \pi \delv(t) \bar\rho(t)} \right]^{1/3}\,
\label{rad}
\end{equation} 
resulting from the identification of haloes as connected regions with
overdensity $\delv(t)$ relative to the mean cosmic density
$\bar\rho(t)$, it is clear from equation (\ref{vir}) that $\rho(r)$
depends on the three independent quantities $M$, $E$ and $P$.

Moreover, as we will see in \S~\ref{model}, relaxed haloes grow
inside-out between major mergers, which allows one to express $P$ in
terms of the instantaneous mass and energy accretion rates. Indeed,
under these circumstances their structure within the instantaneous
virial radius at any time $t$ during smooth accretion remains frozen.
Then the definition of the total mass (eq.~[\ref{mass}] with $r$ equal
to $R$) and the virial relation (eq.~[\ref{vir}]) lead to
\begin{equation}
M(t)-M_0=\int_{R_0}^{R(t)} \der r\,4 \pi r^2\,\rho(r)
\label{mt}
\end{equation}
\begin{eqnarray}
2E(t)-2E_0=-\int_{R_0}^{R(t)} 4 \pi
r^2\,\rho(r)\,\frac{GM(r)}{r}\,\der r \nonumber \\
+\, 4\pi\left[R^3(t)P(t)-R_0^3P_0\right]\,,{~~~}
\label{ener}
\end{eqnarray}
where $R(t)$, $M(t)$, $E(t)$ and $P(t)$ are the virial radius, mass,
energy and confining pressure of the evolving halo at $t$ and $R_0$,
$M_0$, $E_0$ and $P_0$ their respective values at an arbitrary initial
epoch $t_0$ after the last major merger. By differentiating these two
equations and taking into account the Jeans equation $\der P/\der
r=-GM(r)\rho(r)/r^2$, we obtain the relations
\begin{equation}
\dot M= 4 \pi R^2(t)\,\rho[R(t)]\,\dot R
\label{dotM}
\end{equation}
\begin{equation}
\dot E=\dot M\left\{-\frac{GM(t)}{R(t)}+
\frac{3 P(t)}{2\rho[R(t)]}\right\}\,,
\label{dotE}
\end{equation}
from which one can infer, aside from the inside-out evolving density
profile (see MRSSS and \citealt{Hio03}), the pressure at the
instantaneous virial radius
\begin{equation}
P(t)=\frac{\dot M}{6\pi R^2(t)\dot R}\left[\frac{\dot E}{\dot M}+
\frac{GM(t)}{R(t)}\right]\,,
\label{P}
\end{equation}
where $\dot R$ is a known function of $M$, $\dot M$ and $t$ following
from the differentiation of equation (\ref{rad}) with $R$ and $M$ some
functions of $t$.

Thus, the density profile $\rho(r)$ of relaxed haloes at $t$ depends
on the four independent quantities: $M$, $E$, $\dot M$ and $\dot E$,
fixing the right hand-side member of equation (\ref{vir}). Of course,
these are not the only variables on which $\rho(r)$ may depend since
the uniqueness of the solution of equation (\ref{vir}), regarded as an
integral equation for $\rho(r)$, is not warranted in general (not even
in the case that $\rho(r)$ can be restricted, once $M$ is fixed, to a
one-parameter function). Therefore, $\rho(r)$ may in principle depend
on other quantities, perhaps related to some aspects of the
aggregation history of the halo as the correlations mentioned in
\S~\ref{intro} suggest. In Appendix \ref{general}, we show however
that for density profiles of the NFW-like universal form the solution
of equation (\ref{vir}) is unique. Consequently, the only parameter
required to fix the density profile in this case, $\rs$, must depend
exclusively on $M$, $E$, $\dot M$ and $\dot E$. In other words,
\emph{all relaxed (spherical, non-rotating) haloes with identical
current values for these four independent quantities necessarily have
the same value of $\rs$ regardless of their individual aggregation
history}.

A first implication of this result is that it allows one to understand
the origin of the mass-concentration relationship shown by haloes at
any given epoch without the need to presume any dependence of $\rs$ on
the time of their formation. As we have just shown, $\rs$ (or $c$)
depends, at a given $t$, on $M$, $E$, $\dot M$ and $\dot E$, while the
latter three quantities are expected to correlate, in any given
cosmology, with $M$. To see this latter point we must take into
account that the kinetic energy of protohaloes at some early enough,
otherwise arbitrary, time $\ti$ arises essentially from the Hubble
flow at that epoch, whereas their potential energy should not be far
from that of a peak of density contrast
\begin{equation}
\delta_{\rm i}=\delta_{\rm c}\frac{D(\ti)}{D(t)}\,
\label{deltat}
\end{equation}
on the scale
\begin{equation}
R_{\rm i}=\left[\frac{3M}{4\pi\,\bar\rho(\ti)}\right]^{1/3}\,,
\label{radi}
\end{equation} 
with the most probable density profile \citep{BBKS}. In equation
(\ref{deltat}), $\delta_{\rm c}$ is the linearly extrapolated density
contrast for spherical collapse at the present time (equal to 1.69 in an
Einstein-de Sitter universe) and $D(t)$ the, also cosmology-dependent,
linear growth factor. Thus, once the arbitrary time $\ti$ is fixed,
equations (\ref{deltat}) and (\ref{radi}) can be used to estimate the
(conserved) total energy $E$ of haloes as a function of $M$ and $t$.
On the other hand, the mass accretion rate $\dot M$ of haloes at $t$
is well approximated by (Raig, Gonz\'alez-Casado, \& Salvador-Sol\'e
2001) 
\begin{equation}
\ra(M,t)=\int_0^{\delm M} \Delta M\, r\PS(M,\Delta M,t)\,\der \Delta M\,,
\label{ra}
\end{equation}
where $r\PS(M,\Delta M,t)$ is the usual Extended Press-Schechter (EPS)
merger rate \citep{LC93} and $\delm$ is the fractional mass increase
above which a merger is considered major and, therefore, does not
contribute to accretion anymore (Salvador-Sol\'e et
al.~1998). Finally, by introducing the former two functions, $E(M,t)$
and $\dot M(M,t)$, into the fundamental relationship
\begin{equation}
\dot E(M,t)= \frac{\partial E}{\partial M}\,\dot M+
\frac{\partial E}{\partial t}\,,
\label{dotE2}
\end{equation}
one can also estimate the energy accretion rate $\dot E$ in terms of
$M$ and $t$.

We want to stress that this interpretation leads to quantitative
predictions of the mass-concentration relation from the EPS theory
that are, except for rare very massive haloes, in good agreement with
the results of \nbody simulations for different cosmologies and
redshifts (MRSSS; \citealt{Hio03}).

\section{A PHYSICAL MODEL FOR THE EVOLUTION OF $\rss$}
\label{model}

We have just seen that the mass-concentration relation can be explained
without presuming any dependence of the density profile of haloes on
their mass aggregation history. In \S\ \ref{sim}, it will be shown
that the same is true with regards the correlations reported by WBPKD and
ZMJB. But to do this we need first to determine how $\rs$ evolves
along the main branch of halo merger trees.

In \S~\ref{sradius} it was assumed that, during the gentle accretion
phase between major mergers, relaxed haloes grow inside-out, i.e.,
their inner structure remains frozen. This implies that haloes
evolving along typical accretion tracks, represented by the solution
of the differential equation (Raig et al.~2001)
\begin{equation}
\frac{\der M}{\der t}=\ra[M(t),t]\,,
\label{mat}
\end{equation}
where $\ra(M,t)$ is the mass accretion rate given by equation
(\ref{ra}), maintain an unaltered value of $\rs$ or, equivalently, that
such typical mass accretion tracks coincide with typical
$\rs$-isopleths. Moreover, according to the result of
\S~\ref{sradius}, these $\rs$-isopleths do not depend on the past
evolution of haloes but just on their current mass. Thus, haloes
undergoing a major merger will jump from the $\rs$-isopleth they were
initially following to a new $\rs$-isopleth according only to their
change in mass, with the constant value of $\rs$ associated with each
accretion track given by the mass-concentration relation at any
arbitrary redshift. Therefore, the typical value of $\rs$ for a halo
evolving along the main branch of a merger tree is simply that
corresponding to the typical $\rs$-isopleth (or mass accretion
track) that is intersected, at each moment, by the Mass Aggregation
Track (MAT) followed by the halo.

At this stage, it is important to remark that the key assumption in
this scheme, namely that haloes grow inside-out between major mergers,
is fully supported by the results of numerical simulations. When a
halo evolves through accretion and major mergers, $M$ necessarily
increases but $\rs$ may not do so; in periods of inside-out growth,
$\rs$ will be kept unaltered. The opposite is not true, of course: the
invariance of $\rs$ does not guarantee that the inner structure of the
halo remains frozen as the characteristic density $\roc$ may
vary. What should certainly be a secure signature for the inside-out
growth of haloes along the main branch of merger trees is the
simultaneous constancy of both the scale radius $\rs$ and the mass
$M\s$ interior to it, as these two quantities fix the two degrees of
freedom of the NFW-like density profile. In this case, the inner
structure remains necessarily frozen and the mass $M$ grows just
because $R$ does. Therefore, the inside-out growth of haloes between
major mergers must lead to values of $\rs$ and $M\s$ that are kept
simultaneously constant along the main branch of halo merger trees
between any two sudden increases of $M$ marking major mergers. As
Figures 2 and 3 of ZMJB illustrate, this schematic behaviour is indeed
found in numerical simulations (see also our Figs.~\ref{5} and \ref{6}
below where we depict the results of Monte Carlo simulations built
under the assumption of halo inside-out growth during the accretion
phases). It is not inconsistent either with the results obtained by
\citet{Gea04} from numerical simulations of cluster-sized haloes
showing that most of the matter that ends up, at $z=0$, near the
centre is not at the centre of their main progenitor at $z=4$. The
reason is that the inside-out growth only holds between major mergers,
while the last major merger where current cluster-sized haloes were
rearranged took place at a redshift much smaller than 4 (typically at
$z\approx 1$).

\begin{figure}
\centerline{\includegraphics*[width=9.5cm]{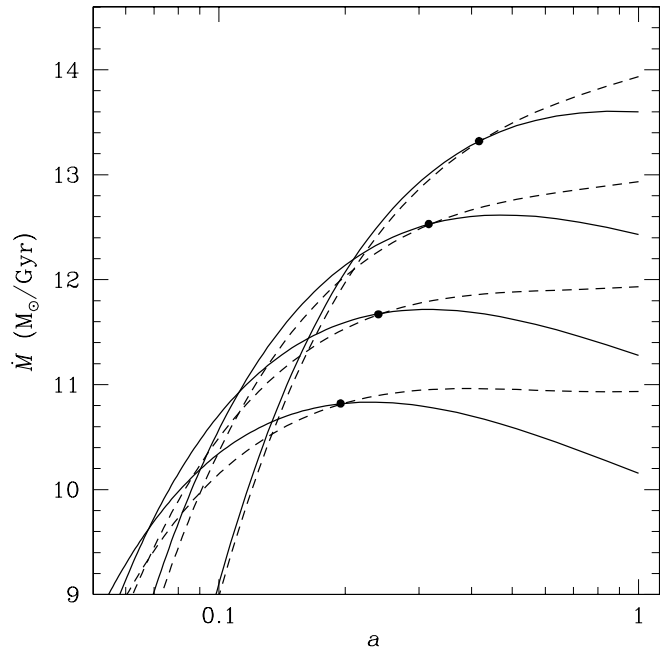}}
\caption{Typical mass accretion rates (solid lines) compared with the
quantity $1.2\,H(t)\,M(t)$ (dashed lines) found along the typical main
branch of halo merger trees, here referred to as MATs, leading at
$z=0$ to halo masses equal to $10^{12}$, $10^{13}$, $10^{14}$, and
$10^{15}$ M$_\odot$ (from bottom to top) for the same cosmology used
by ZMJB and WBPKD. The typical MAT corresponding to each halo mass is
drawn from equation (18) for the appropriate value of $\alpha$
according to WBPKD. The epoch during which each pair of curves
intersects (big dots) gives the characteristic redshift $z\fa$
separating the fast and slow accretion regimes in the corresponding
MATs.}
\label{1}
\end{figure}

{}From Figures 2 and 3 of ZMJB (and Figs. \ref{5} and \ref{6} below)
it is also apparent that the phases of inside-out growth are only
found at relatively low redshifts. There is nothing strange in this
behaviour. The inside-out growth of a halo during spherical infall is
strictly warranted provided only that the orbital period of particles
in any given shell is much smaller than the characteristic time of
mass infall (see MRSSS and references therein) or, equivalently, that
accretion is slow enough. In periods of fast accretion, shell-crossing
is expected rather to induce the violent relaxation of the halo just
as major mergers do. As one goes back in time, the orbital period of
particles in the outermost shell becomes of the order of, and even
larger than the typical infall time of accreted matter in the
cosmology considered, so the conditions for inside-out growth no
longer hold. The redshift $z\fa$ at which the transition between fast
and slow accretion regimes typically occurs can thus be estimated by
comparing the orbital period of particles at the virial radius of the
relaxed halo,
\begin{equation}
\tau=\left[\frac{4}{3}\pi\,G\,\delv(t)\,\bar \rho(t)\right]^{-1/2}\,,
\label{tau}
\end{equation}
with the characteristic time of mass infall during accretion, equal to
$M/\ra(M,t)$. Thus, given the equality $\bar \rho(t)=3
H^2(t)\Omega\m(t)/(8\pi G)$, where $H(t)$ and $\Omega\m(t)$ are
respectively the Hubble and density parameters at $t$, and that, in
the relevant range of cosmic times, $\Omega\m(t)$ and $\delv(t)$ can
be respectively approximated by 1 and 200 (for $\delv(t)$ according to
\citealt{bn98}, for which $\delm\sim 0.21$; \citealt{Hio03}), the
condition for slow accretion adopts the form
\begin{equation}
\ra[M(t),t]\ll 10\, H(t)\,M(t)\,.
\label{condition}
\end{equation}
In Figure \ref{1}, we show the redshifts $z\fa$ at which the typical
mass accretion rates encountered along MATs corresponding to different
halo masses satisfy the equality $\ra[M(t),t]=1.2H(t)\,M(t)$, where
the factor 1.2 has been chosen to give the best agreement with the
results of ZMJB (see \S~\ref{zetal}).

\begin{figure}
\centerline{\includegraphics*[width=9.5cm]{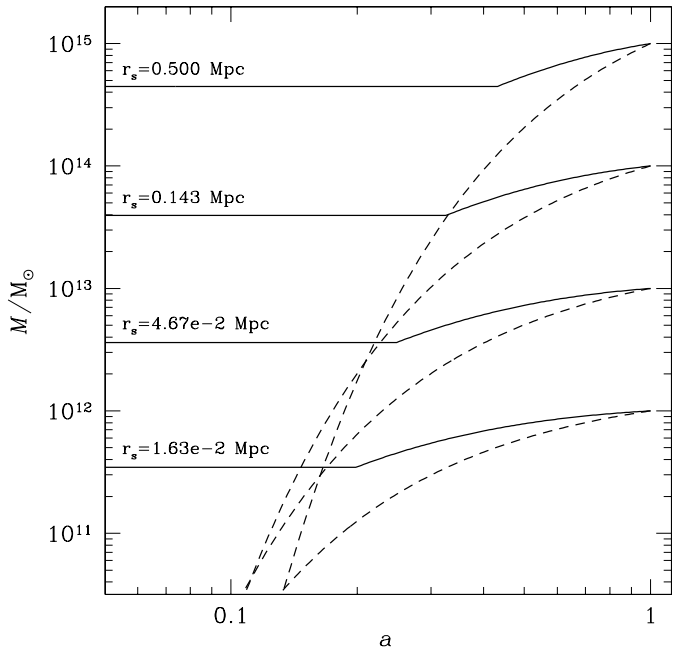}}
\caption{Typical $\rs$ (and $M\s$) isopleths (solid lines) for haloes
with the same final masses (from bottom to top) as in Figure 1. The
$\rs$ values associated with the different curves have been calculated
from the mass-concentration relation at $z=0$ given by
\citet{Bull01}. (The corresponding $M\s$ values are given by
eq. [\ref{msm}].) The curved part of each isopleth corresponds to the
slow accretion regime, while the horizontal part corresponds to fast
accretion. Dashed curves show the typical MATs of haloes for the same
final masses.}
\label{2}
\end{figure}

According to the previous discussion, the typical $\rs$-isopleths
strictly correspond to the typical evolutionary tracks followed by
haloes during \emph{slow} accretion. That is, to infer them we must
solve equation (\ref{mat}) taking into account only the slow
contribution to the accretion rate. Therefore, for $z\le z\fa$ where
accretion is typically slow, the typical $\rs$-isopleths will coincide
with the solutions of equation (\ref{mat}) based on equation
(\ref{ra}), while, for $z> z\fa$ where accretion is typically fast and
the contribution of slow accretion to the right hand-side member of
equation (\ref{mat}) is null, they will be kept frozen to the values
reached at $z\fa$ (see Fig. \ref{2}). In this manner, our model for
the typical evolution of $\rs$ along MATs becomes valid \emph{at all
redshifts}. Note that, compared to other prescriptions for the
evolution of $\rs$ available in the literature, our model stands out
for being physically motivated and for eliminating the requirement of
knowing (and, hence, of storing) the individual aggregation histories
of haloes in order to infer the typical value of their $\rs$ at any
moment since this is estimated from their current mass only\footnote{A
FORTRAN programme that computes the typical $\rs$ value for relaxed
haloes in CDM cosmologies according to our prescription can be
downloaded from {\tt http://www.am.ub.es/cosmo/gravitation.htm}}.

\begin{figure}
\centerline{\includegraphics*[width=12cm]{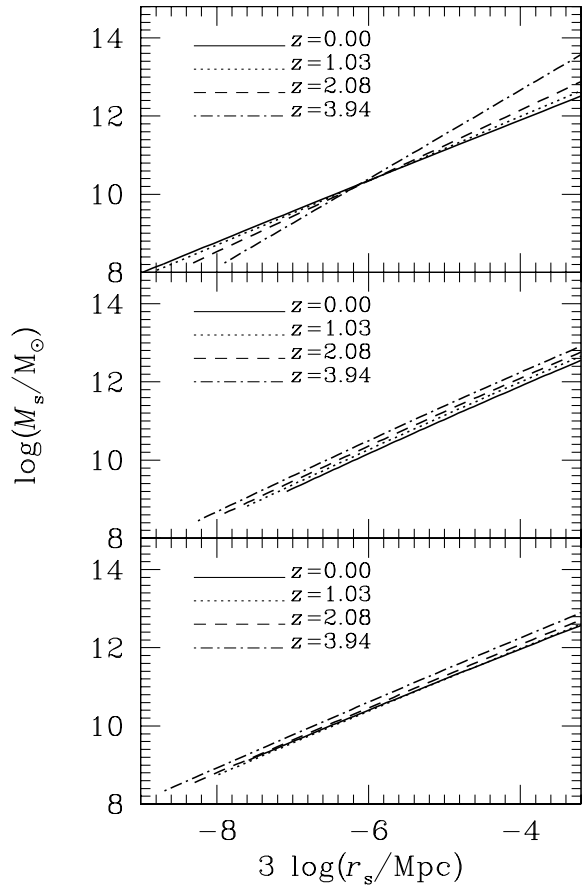}}
\vskip .5cm
\caption{Typical $\rs - M\s$ relations obtained from the
mass-concentration relations at different redshifts drawn from Zhao et
al.~(2003b) data (top) and from the toy models provided by Eke et
al.~(2001) (center) and \citet{Bull01} (bottom) for the same
$\Lambda$CDM cosmology as in Zhao et al.~(2003b), identical to that
used throughout the rest of the paper (and in WBKPD and ZMJB studies
as well) but for the normalisation constant $\sigma_8$, here equal to
0.9.}
\label{3}
\end{figure}

Given that the inner structure of haloes, determined by the values of
$\rs$ and $M\s$, is kept unaltered along slow mass accretion tracks,
the typical $\rs$-isopleths coincide, of course, with the typical
$M\s$-isopleths. The constant value of $M\s$ associated to an
$\rs$-isopleth can be obtained from the constant value of this latter
parameter and the mass of the halo at any arbitrary time along that
track from the relation
\begin{equation}
M\s=\frac{(\ln 2- 1/2)M}{\ln[1+R(M)/\rs]-[R(M)/\rs]/[1+R(M)/\rs]}\,,
\label{msm}
\end{equation}
arising from the NFW universal density profile, with $R(M)$ given by
equation (\ref{rad}). (Hereafter we assume eq.~[1] with $\beta=1$;
were $\beta$ different from unity, eq. [\ref{msm}] should be replaced
by the corresponding expression.) This coincidence between the $\rs$-
and $M\s$-isopleths implied by our model leads to an interesting
prediction that can be easily checked. The relation between $\rs$ and
$M\s$ along the different isopleths defines an $\rs - M\s$
relationship that should be typically satisfied by haloes regardless
of their mass and redshift. Such a relationship can be obtained from
equation (\ref{msm}) and the mass-concentration relation {\it at any
arbitrary redshift\/}. Therefore, we are led to the conclusion that
the $\rs - M\s$ relations inferred from the mass-concentration
relations at different redshifts must overlap. (Note that this
conclusion is independent of the form of the $\rs$- or $M\s$-isopleths
and the exact value of $\beta$.) In Figure \ref{3}, we plot the $\rs -
M\s$ relations that result from the empirical mass-concentration
relations obtained by Zhao et al.~(2003b) at different redshifts, as
well as those calculated, for the same redshifts and $\Lambda$CDM
cosmology, from the toy-models provided by \citet{Bull01} and Eke et
al.~(2001). In the latter two cases, the $\rs - M\s$ relations at
different redshifts are all very similar, indeed, confirming our
expectations. There is only a slight shift with $z$ which can be
attributed to the inherent inaccuracy of the toy-models. However, in
the case of Zhao et al., the $\rs - M\s$ relations show a substantial
$z$-dependence. This discrepancy between the behaviours of the $\rs -
M\s$ relations derived from the mass-concentration relations provided
by different authors is not unexpected given the notable deviation of
the mass-concentration relations of Zhao et al. from those of the
other two groups at high $z$ (see Fig.~2 of Zhao et al.~2003b).  The
origin of such a deviation is unclear and deserves further
investigation. We note however that the Zhao et al. mass-concentration
relations traced by the points drawn from their 25 $h^{-1}$ Mpc size
box simulation, with softening length equal to 2.5 $h^{-1}$ Kpc, closely
follow those of the other two groups inferred from simulations with
similar softening lengths (equal to 1.8 and 0.4 $h^{-1}$ Kpc in
Bullock et al. and Eke et al., respectively), while the discrepancy
begins to be apparent and becomes definitely marked when the Zhao et
al. mass-concentration relations are traced by the points drawn from
their 100 and 300 $h^{-1}$ Mpc size box simulations, with softening
lengths respectively equal to 10 and 30 $h^{-1}$ Kpc.

\section{REINTERPRETING SOME EMPIRICAL CORRELATIONS}\label{sim}

\begin{figure*}
\vskip -3.5 cm
\centerline{\includegraphics*[width=14cm]{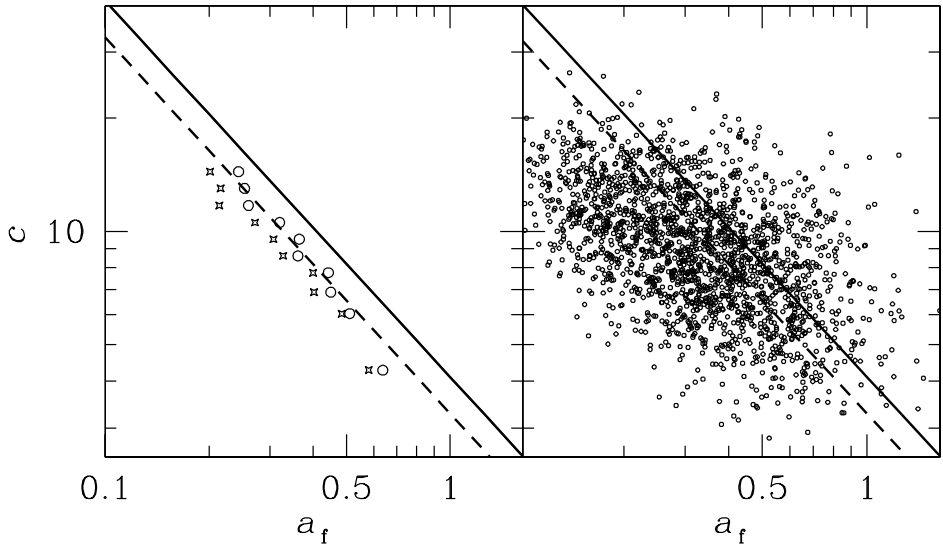}}
\vskip -3 cm
\caption{The $c - a(z\f)$ correlation obtained from 2000 Monte Carlo
MATs derived in the framework of the EPS formalism using our physical
model for the evolution of $\rs$ for haloes with masses (at $z=0$)
larger than $1.43\times 10^{12}$ M$_\odot$. The quantity $a\f$
plotted in the right panel is the normalized cosmic expansion factor,
$a(z\f)/a(0)$, corresponding to the redshift of formation, $z\f$, of
each halo obtained in the way prescribed by WBKPD (see text) from its
individual MAT, whereas in the left panel it is the typical value
(mean identified by circles and median by crosses) for haloes in final
mass bins of 200 objects. The straight lines show the best linear fit
to the correlation obtained from \nbody simulations by WBPKD (solid
line) as well as its shift (dashed line) so to fit the correlation
obtained by the same authors from Monte Carlo simulations using their
toy-model for $\rs$ that assumes some dependence of this parameter on
the individual halo MAT.}
\label{4}
\end{figure*}

\subsection{The $\bmath{c}$ vs. $\balpha$ correlation}\label{ca}

WBPKD showed that the MAT\footnote{WBPKD did not distinguish between
minor and major mergers and used the word accretion for what is here
referred to as aggregation.} of individual haloes of mass $M_0$ at a
redshift $z_0$ (as well as the average MAT of haloes with $M_0$ at
$z_0$) is typically well fitted by the function
\begin{equation}
M(z)=M_0 \exp\left\{-\alpha \left[\frac{a(z_0)}{a(z)}-1\right]\right\}\,,
\end{equation}
with $a(z)$ the cosmic scale factor at the redshift $z$. Thus, by
defining the redshift $z\f$ of formation of haloes as the epoch in
which the logarithmic slope of their respective MAT takes some
fiducial value $S$ ($\sim 2$), the previous result implies the following
relation between $z\f$ and $\alpha$
\begin{equation}
\alpha=\frac{a(z\f)}{a(z_0)}\,S\,.
\label{alpha}
\end{equation}
In addition, these authors found that the concentration $c$ of haloes
at a given redshift correlates with their redshift of formation or,
equivalently, with the parameter $\alpha$ associated to their
individual MAT. From this result, WBKPD concluded that the density
profile of haloes is determined by their aggregation history.

There is a different possibility, however. As shown in
\S~\ref{sradius}, mass and energy conservation, together with the fact
that, in any given cosmology, the typical accretion rate of haloes is
a function of $M$ and $t$, allows one to relate the concentration $c$
of haloes at some given redshift to their mass. This implies, in turn,
that $c$ also correlates with any arbitrary function of $M$, in
particular with $\alpha(M)$ giving the typical (mean or median) value
of $\alpha$ for haloes of mass $M$. Since the individual values of 
$\alpha$ obviously scatter around their typical one, the $c -
\alpha(M)$ correlation arising from the mass-concentration relation
will induce a secondary correlation between $c$ and the $\alpha$ value
corresponding to each MAT or, equivalently, with its associated
$a(z\f)$ value. In other words, the $c - a(z\f)$ correlation reported
by WBPKD might simply be a consequence of the well-known
mass-concentration relation and, just like it, might not imply any
dependence of $\rs$ on the aggregation history of haloes.

To check the validity of this alternative explanation we ran 2000
Monte Carlo MATs for present-day haloes with masses according to the
\citet{st99} mass function and satisfying the same restriction,
$M>1.43\times 10^{12}$ M$_\odot$, as in WBPKD and followed, in each,
the evolution of $\rs$ predicted by the model developed in
\S~\ref{model} that does not presume any dependence of this parameter
on the halo aggregation history. These Monte Carlo MATs were built
from the conditional probability that haloes of a given mass at a
given epoch arise from smaller objects at earlier epochs given by the
EPS theory \citep{LC93}. The time steps adopted were small enough to
ensure the inclusion of only one merger with $\Delta M$ above the mass
resolution of the simulations according to \citet{LC93}. In this way,
to be sure that the inferred MATs really correspond to the main branch
of merger trees, we only had to impose that the relative mass increase
in any time step was smaller than unity. The values of $\rs$
corresponding to the different isopleths in our physical model for the
evolution of this parameter were drawn from the mass-concentration
relation at $z=0$ of \citet{Bull01}.

In Figure \ref{4}, we show the $c - a(z\f)$ correlation at $z=0$
obtained from these Monte Carlo simulations. To compensate for the
fact that our model for $\rs$ does not include random deviations of
$E$, $\dot M$ and $\dot E$ from their typical values that depend
exclusively on $M$ and $t$, we have added, according to \citet{ji00},
Gaussian deviates in log($c$) with a dispersion equal to $0.09$.  The
main trend of this correlation reproduces satisfactorily that found in
WBPKD's $N$-body simulation (see their Fig. 6). The only small
differences concern the slightly larger scatter and the small shift (a
little more marked in the medians than in the means) in the Monte
Carlo correlation relative to the \nbody one. This behaviour is
identical to that shown by the correlation obtained by WBKPD in the
Monte Carlo simulation they performed, using a slightly different
implementation of the EPS theory, to check the validity of their
toy-model for $\rs$ that includes an explicit dependence of this
parameter on the individual halo MAT. These authors found, indeed, a
scatter systematically larger (by 25\%) in the Monte Carlo correlation
than in the \nbody counterpart and the correlation itself shifted
exactly by the same amount as found in the present work. As mentioned
by these authors, the small deviations of the Monte Carlo predictions
with respect to the \nbody results are likely caused by the well-known
inaccuracies of the EPS formalism (and its practical implementation;
see e.g. \citealt{SK99} and Raig et al. 2001). In any case, what is
important here is that our physical model for $\rs$ is capable of
recovering the correlation found by WBPKD at the same level as their
toy-model despite the fact that, unlike their model, ours does not
presume any dependence of $\rs$ on the halo aggregation history.

\subsection{The $\bmath{M_{\rm s}}$ vs. $\bmath{r_{\rm s}}$ correlation}
\label{zetal}

ZMJB recently found that haloes show a tight correlation between the
$\rs$ and $M\s$ values scaled to those of the main progenitor at a
redshift $z\tp$ where the MAT shows a characteristic change of
slope\footnote{The discontinuity in the MAT slope coincides with a
maximum or ``turning-point'' in the function $\log(\vc)+0.25
\log[H(z)]$, where $\vc$ is the circular velocity of the halo along
its MAT, which provides a well-defined practical procedure to
determine $z\tp$. Note also that, like WBPKD, ZMJB used the word
accretion for what is here referred to as aggregation.}, which they
interpreted as an indication of the passage from fast to slow
aggregation regime. Moreover, they showed that the log-log slope of
this scaled $\rs - M\s$ correlation changes depending on whether the
halo redshift is larger or smaller than the respective $z\tp$
value. This finding was interpreted as evidence that the inner
structure of haloes depends on their aggregation history. Still, ZMJB
did not offer any explanation neither for the origin of this
correlation nor for its dependence on the halo redshift relative to
$z\tp$.

In fact, the frequency of major mergers cannot be held responsible for
the change of slope in the MAT of a halo at $z\tp$ because, in the
$\Lambda$CDM cosmology, major mergers are, for the present-day halo
masses considered by ZMJB ($\sim 10^{13}$ M$_\odot$), even more
frequent in the slow than in the fast aggregation regimes defined
according to the heuristic redshift $z\tp$. What could explain,
instead, the characteristic change of slope in MATs observed at that
redshift is the transition from fast to slow \emph{accretion} at
$z\fa$ mentioned in \S\ \ref{model}. Besides, the correlation between
$\rs$ and $M\s$ predicted by the model developed in that section is
expected to split into two tight correlations similar to those found
by ZMJB when the values of the two quantities are scaled to those of
the main progenitor at $z\fa$. Note that the $\rs - M\s$ relation
shown in Figure \ref{3} is very nearly linear in log-log. Therefore,
the fact that major mergers and fast accretion tend to increase with
decreasing redshift the values of $\rs$ and $M\s$ as haloes evolve
along MATs plus the slight $z$-dependence of the slope of the log-log
$\rs - M\s$ correlation derived from the Zhao et al. (2003b)
mass-concentration relations (see Fig. \ref{3}) will translate into a
small bending (on top of the slight intrinsic one; see Appendix B) of
the log-log $\rs - M\s$ correlation shown by haloes along
MATs. Consequently, by scaling the values of $\rs$ and $M\s$ to those
of the main progenitor at $z\tp$ (or $z\fa$) along each individual
MAT, such a bent correlation will split into two essentially linear
relations on both sides of that redshift, with that corresponding to
the fast regime being slightly steeper as found by ZMJB. Note however
that, according to our model, since the values of $\rs$ and $M\s$ do
not depend on the particular MAT the haloes have followed, the scaled $\rs
- M\s$ correlations reported by ZMJB should not rely on their scaling
by the specific values of $\rs$ and $M\s$ corresponding to the main
progenitor found at $z\tp$ (or $z\fa$) along each {\it particular\/}
MAT. Using as reference the object found at $z\tp$ (or $z\fa$) along
the {\it typical\/} MAT for haloes with the mass of that being studied
should yield essentially the same results.

The validity of our interpretation of the ZMJB findings can be
checked, like in the case of the WBPKD correlation, by means of Monte
Carlo simulations. Unfortunately, this is not so immediate in the
present case. As explained in \S~\ref{model}, our model for the
evolution of $\rs$ (and $M\s$) is not fully consistent with the
mass-concentration relations found by Zhao et al. at high
redshifts. However, we can still check it on the scaled $\rs - M\s$
correlation arising from the mass-concentration relations of
\citet{Bull01} or Eke et al.~(2001) that do not suffer from this
problem. As, in the latter case, the general $\rs - M\s$ correlation
does not depend on $z$, we do not expect its $z$-induced bending along
MATs. But, as shown in the Appendix \ref{correlation}, this general
correlation still has some intrinsic, less marked, convexity, which
will also make it split, when properly scaled, into two linear
correlations similar to those found by ZMJB, with the reference halo
in the individual or typical MATs playing exactly the same role.

\begin{figure*}
\centerline{\includegraphics*[width=15cm]{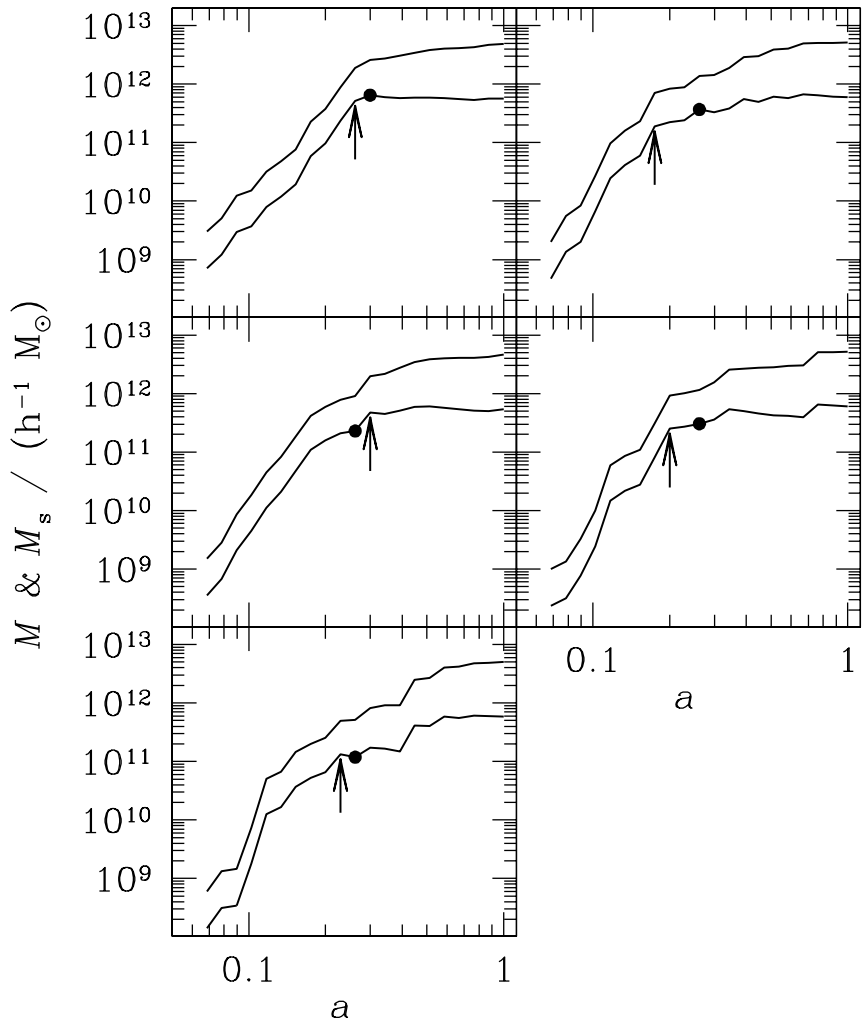}}
\vskip -.7cm
\caption{Monte Carlo realisations (upper curves) of MATs and the
corresponding $M\s$ tracks (lower curves) derived from the values of
$\rs$ plotted in Figure \ref{6} for the same cosmology and similar
final halo masses as in ZMJB. The location of the redshift $z\fa$ (big
dots) marking the typical frontier between fast and slow accretion
regimes is compared to the redshift $z\tp$ (indicated by arrows)
marking the characteristic knee in the mass aggregation tracks
indicated by ZMJB. To compare with Figure 2 of ZMJB.}
\label{5}
\end{figure*}

\begin{figure*}
\centerline{\includegraphics*[width=15cm]{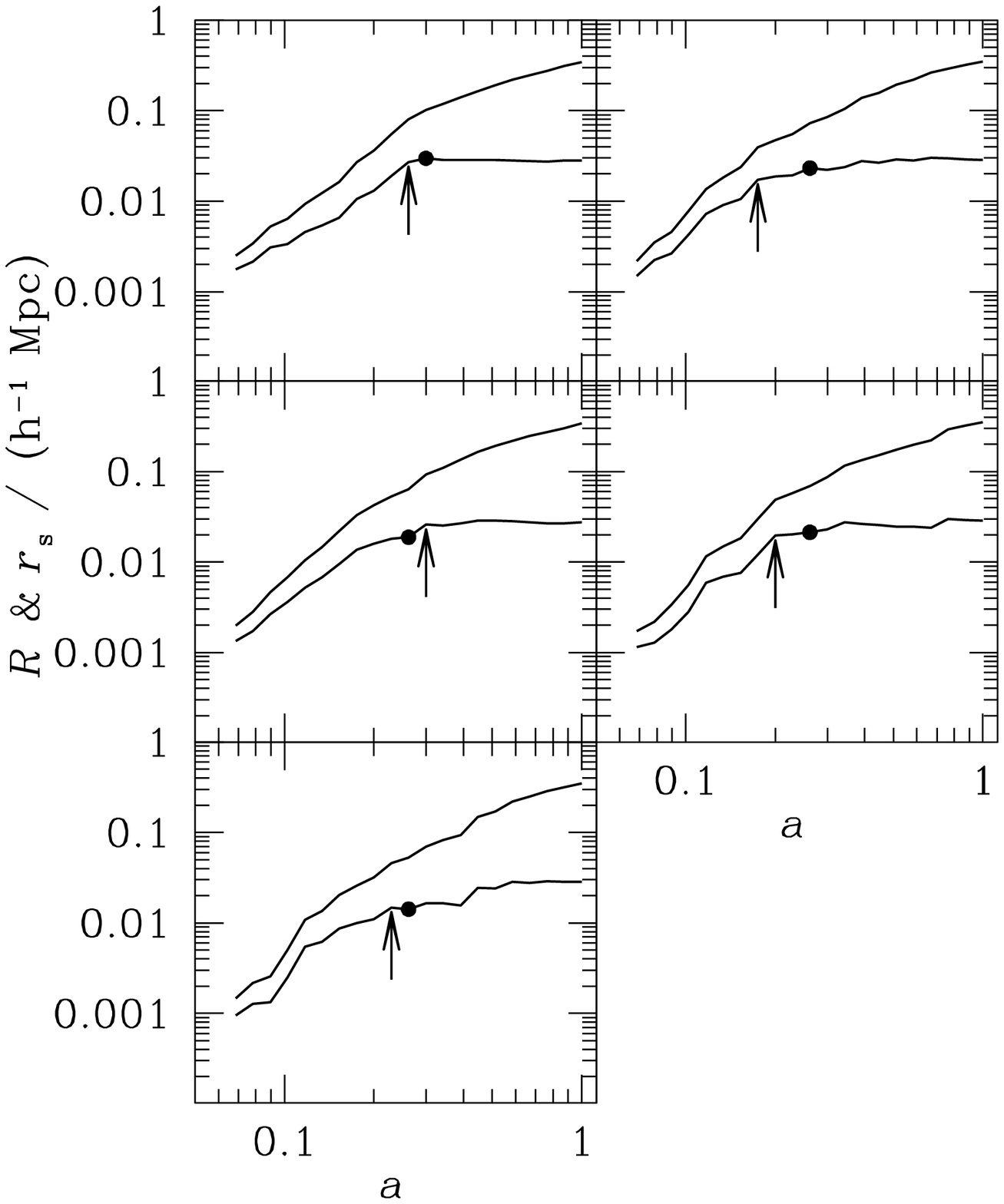}}
\vskip -.7cm
\caption{Evolving virial radii $R$ (upper curves) and scale radii
$\rs$ (lower curves) predicted by our physical model for the evolution
of this parameter corresponding to the Monte Carlo MATs plotted in
Figure \ref{5}. To compare with Figure 3 of ZMJB.}
\label{6}
\end{figure*}

With this aim, we performed Monte Carlo MATs as explained in
\S~\ref{ca} but now following also the evolution of $M\s$. In Figures
\ref{5} and \ref{6} we show a few of these realisations for final halo
masses similar to those appearing in Figures 2 and 3 of ZMJB. To
better mimic the ZMJB results, the mass increases produced along the
MATs were accumulated between each couple of consecutive output times
chosen identical to those in the ZMJB \nbody simulations. The $z\tp$
values, marked on these tracks by arrows, were obtained following the
prescription given by ZMJB, whereas the $z\fa$ values, marked by big
dots, were calculated using the same criterion as in Figure
\ref{1}. Despite the fact that our $\rs$ and $M\s$ tracks do not
include the effects of the random deviations of $E$, $\dot M$ and
$\dot E$ from their respective typical values, they show a remarkable
similarity to those depicted in Figures 2 and 3 of ZMJB. There is good
agreement not only in the general shape of the MATs and the associated
$\rs$ and $M\s$ tracks (see, in particular, the effects of inside-out
growth during slow accretion), but also in the typical location of the
$z\tp$ values. From these figures we can see that every $z\tp$ is
close to the corresponding $z\fa$ marking the change of accretion
regime along each MAT, which confirms our guess on the origin of the
change in slope of MATs. It is interesting to note that the procedure
followed to determine $z\tp$ tends to locate this characteristic
redshift just after a major merger or an interval of fast
accretion. This introduces a systematic discontinuity in the slope of
the MATs at that redshift that explains the two-branch analytical
fitting formula proposed by Zhao et al. (2003b) for the average MAT of
haloes. However, when $z\tp$ is not privileged as in WBPKD, the
typical MAT of haloes of mass $M$ shows no such discontinuity.

\begin{figure*}
\centerline{\includegraphics*[width=15cm]{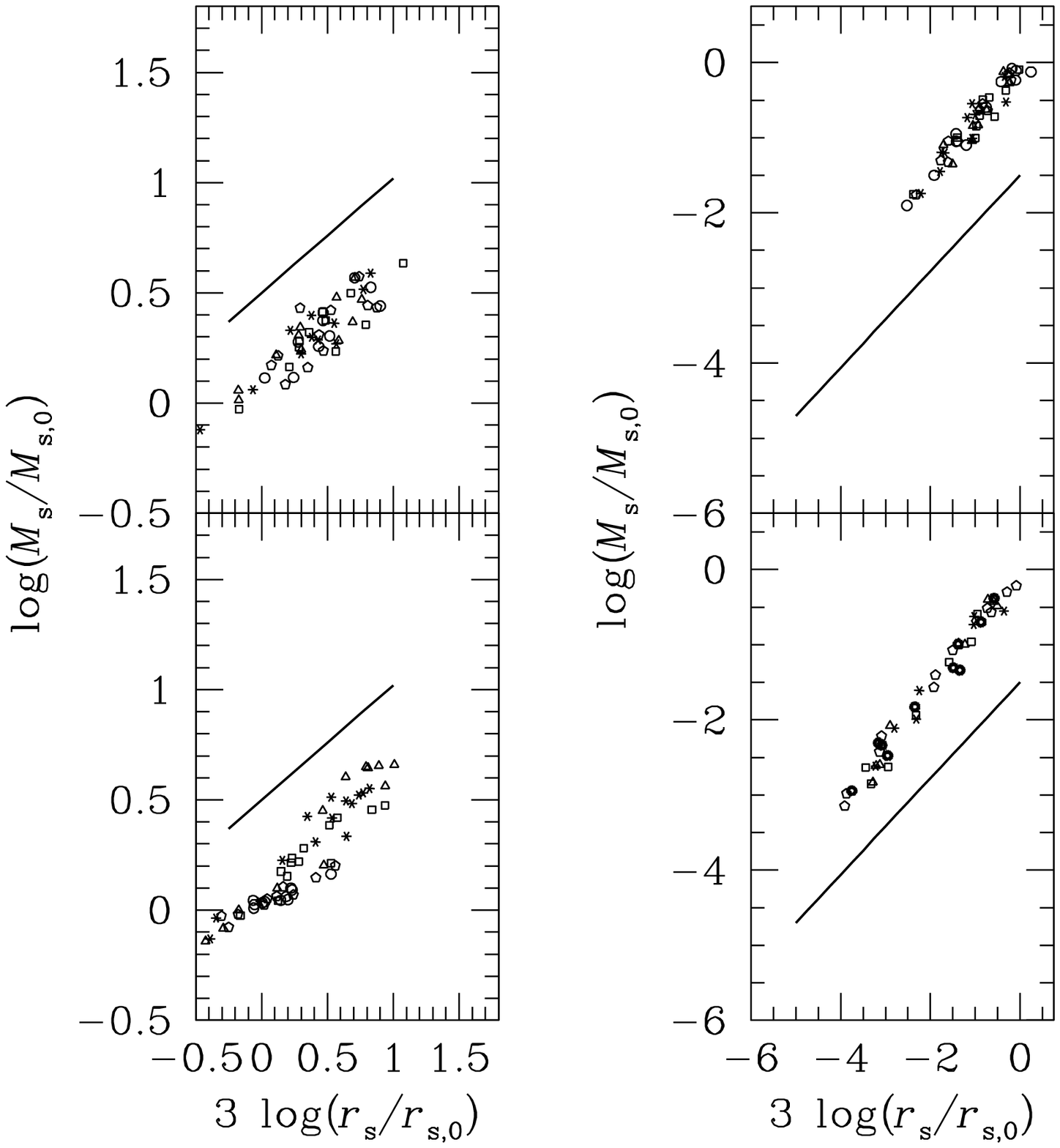}}
\vskip -.5cm
\caption{Scaled $\rs - M\s$ correlations drawn from objects along the
Monte Carlo MATs shown in Figures \ref{5} and \ref{6} at the same
output times of the ZMJB simulations for the slow (left) and fast
(right) accretion regimes. The values $\rso$ and $M\so$ used to scale
the $\rs$ and $M\s$ variables and to separate both aggregation regimes
correspond to the halo found at $z\fa$ along the typical MAT of each
object (top) and to the halo found at $z\tp$ along the real individual
MAT of each object (bottom). Different symbols are used for objects
along different MAT realisations. The straight lines show the slopes
of the similar, though with a different origin (see text), empirical
correlations found by Zhao et al.~(2003b), shifted by the same amount
as in their Figure 7.}
\label{7}
\end{figure*}

In Figure \ref{7}, we plot the scaled $\rs - M\s$ correlations
corresponding to the fast and slow accretion regimes, as in Figure 7
of ZMJB, drawn from objects along the Monte Carlo MATs of Figures
\ref{5} and \ref{6} at the same redshifts as in the ZMJB figure. To
mimic the effects of the random deviations of $E$, $\dot M$ and $\dot
E$ in $N$-body haloes, we have added once again Gaussian deviates in
log($c)$ with a dispersion equal to $0.08$ and $0.12$ in the slow and
fast accretion regimes, respectively. (Unlike WBPKD, who impose no
restriction on their halo sample, ZMJB chose relatively isolated final
haloes. Hence, the dispersion in $c$ of these haloes is expected to be
slightly smaller at $z<z\fa$ than in the case of WBPKD. On the
contrary, the fact that haloes should be typically less relaxed at
$z>z\fa$ should cause a higher dispersion in $c$ there.)  In the top
panels, we show the results obtained when the scaling reference is the
halo located at $z\fa$ along the {\it typical\/} MAT for the mass of
each object, while in the bottom panels we show those obtained by
using, instead, the halo at $z\tp$ along each {\it individual\/} MAT
just like in ZMJB. As expected, both variants show identical trends:
tight linear correlations in log-log with similar slopes and scatters,
strongly resembling those in the original ZMJB scaled
correlations. Specifically, the slopes in the fast and slow accretion
regimes are $0.76\pm 0.03$ and $0.65\pm 0.03$, for the typical-MAT
variant, and $0.79\pm 0.02$ and $0.64\pm 0.03$, for the particular-MAT
one. Since in the former case, the correlations obtained do not
involve, by construction, the MAT of individual haloes, we conclude
that the change in slope of these scaled correlations at $z\tp$ (or
$z\fa$) does not imply the dependence of $\rs$ (and $M\s$) on the
individual halo aggregation history. The same conclusion should
therefore hold for the correlations obtained by ZMJB (with slopes in
the fast and slow regimes respectively equal to 0.64 and 0.48).

\section{SUMMARY AND DISCUSSION}\label{dis}

In the present paper, we have shown that the scale radius $\rs$ of
relaxed, non-rotating, spherically symmetric haloes endowed with the
universal density profile \`a la NFW depends exclusively on the
current values of their mass and energy and the instantaneous
accretion rates of these two quantities. There is therefore no room
for any additional dependence on the particulars of their aggregation
histories. Using a simple, physically motivated model for the
evolution of $\rs$ along the main branch of halo merger trees, we have
demonstrated that our finding that $\rs$ does not depend on the halo
past evolution is fully compatible with the well-known
mass-concentration relation and the correlations recently reported by
WBPKD and ZMJB suggesting the opposite conclusion. We note however
that the correlation found by ZMJB has an origin slightly different
from that predicted within our scheme since it appears to be related
to the particular $z$-dependence of the mass-concentration relation
found by these authors compared to that obtained by other groups. The
present results support the explanation recently provided by MRSSS for
the origin of the universal halo density profile.

Apart from the correlations here investigated, \nbody simulations show
that the density of the local environment also correlates with the
frequency of major mergers occurring along the halo aggregation
histories (the higher the density, the higher this frequency;
\citealt{GKK}) and with the value of $\rs$ (the higher the density,
the smaller $\rs$; \citealt{ji00,Bull01}). Hence, some correlation may
exist between $\rs$ and the frequency of major mergers along the main
branch of individual halo merger trees. As is well-known, in its
original form the EPS formalism does not contain information on the
spatial distribution of haloes, so it is not possible to infer from
our model (unless modified along the lines proposed by
\citealt{MW96} or \citealt{Sea01}) any correlation between $\rs$
and local density from our model. However, given that the environment
of a halo should affect its instantaneous accretion rate (the higher
the density, the larger the typical accretion rate) and that, as we
have demonstrated, the value of $\rs$ depends on this latter quantity,
a correlation between $\rs$ and local density is also expected to hold
within our scheme. For the same reason, the fact that haloes located
in high (low) density environments may have had a higher (lower)
chance of merging with other haloes along their past history is not
inconsistent with the present results either. What these results do
not support is the existence of a {\it direct causal relation\/}
between the frequency of major mergers endured by a halo during its
growth and the final value of its scale radius.

The conclusions of the present study rely on some hypotheses that
deserve some discussion. The least controversial of these is that of
the inside-out growth of haloes between major mergers; as shown in
\S~\ref{model}, it is fully supported by the results of numerical
simulations. In contrast, the assumption that haloes are spherically
symmetric and non-rotating is actually not very realistic. Had we not
adopted these simplifying approximations, the angular momentum, the
asphericity of the system (via the inertia tensor) and the anisotropy
of the local velocity tensor would appear in the full vectorial form
of the virial relation, invalidating the proof in Appendix
\ref{general}. We thus cannot discard the possibility that, under more
realistic assumptions, the inner structure of haloes can preserve some
memory of their aggregation history. We are not concerned here, of
course, about the memory embodied in the final value of any extensive
property, such as the total mass, energy or angular momentum, that are
integrated, and hence fully convolved, over the entire evolutionary
history of the system. The kind of memory that is relevant for the
present discussion is rather that associated with the possible imprint
on the inner structure of haloes of \emph{specific events}, such as
the formation of the halo (whatever its definition), the transition
from fast to slow aggregation regimes or the frequency of major
mergers, having taken place at some point in the past. Certainly, the
elongation and the anisotropy of the velocity tensor of haloes
indicate that the episodes of violent relaxation do not proceed until
completion. But this does not necessarily mean, of course, that the
loss of memory produced in those relaxation processes is so small that
precise information on those specific events can be recovered from the
final properties of the spherically averaged density profile. The
results obtained in the present paper rather suggest, along the lines
of those drawn from \nbody experiments mentioned in \S~\ref{intro}
(\citealt{HJS99, Moore99}; WBPKD), that the density profile of haloes
is largely insensitive to the details of their aggregation history. In
any case, they show that, from the correlations found in numerical
simulations, it cannot be concluded that $\rs$ depends on the halo
aggregation history since, at least under the assumption of spherical
symmetry and null-rotation, they are well explained and quantitatively
recovered without presuming such a dependence.


%% Included in this acknowledgments section are examples of the
%% AASTeX hypertext markup commands. Use \url without the optional [HREF]
%% argument when you want to print the url directly in the text. Otherwise,
%% use either \url or \anchor, with the HREF as the first argument and the
%% text to be printed in the second.

\vspace{0.75cm} \par\noindent
{\bf ACKNOWLEDGEMENTS} \par

\noindent This work was supported by Spanish DGES grant
AYA2003-07468-C03-01. We wish to thank the ZMJB team for kindly
providing the data on their mass-concentration relations and James
Bullock and Julio Navarro for making publicly available their
toy-model codes. We are also grateful to Andrei Doroshkevich and Juan
Uson for useful comments.



\clearpage


\clearpage

\appendix

\section{Uniqueness of the NFW solution of the virial 
relation}\label{general}

Consider two steady spherically symmetric, non-rotating,
self-gravitating systems with identical radius $R$ and total mass $M$.
This obviously warrants that their respective density profiles
$\rho(r)$ and $\rho'(r)$ coincide at least at one radius in the open
interval $(0,R)$. We next prove that if the two systems also have
identical total energy $E$ and are subject to the same external
pressure $P$ then the radius at which both profiles coincide cannot
be unique.

Suppose that the density profiles $\rho(r)$ and $\rho'(r)$ (assumed
with realistic central logarithmic slopes greater than $-2$ to ensure
finite central potentials) do intersect each other at {\it a single
radius\/} $r\eq$ within $(0,R)$. The scalar virial relation (\ref{vir})
allows one to write
\begin{equation}
\int_0^R \der M(r)\,\frac{M(r)}{r}=\int_0^R \der M'(r)\,\frac{M'(r)}{r}\,,
\label{A1}
\end{equation}
where $M(r)$ and $M'(r)$ are the mass profiles associated with
$\rho(r)$ and $\rho'(r)$, respectively. By integrating by parts both
sides of this equality, we obtain
\begin{equation}
\int_0^R \der r\,\frac{M^2(r)}{r^2}=\int_0^R \der r\,\frac{M'^2(r)}{r^2}\,.
\label{A2}
\end{equation}
Rewriting equation (\ref{A2}) in the form
\begin{eqnarray}
\int_0^R \der r\,\left[\frac{M^2(r)}{r^2}-\frac{M'^2(r)}{r^2}\right]=
{~~~~~~~~~~~~~~~~~~~~} \nonumber \\
\int_0^R \der r\,\left[M(r)-M'(r)\right]\,
\left[\frac{M(r)+M'(r)}{r^2}\right]=0
\label{A3}
\end{eqnarray}
and integrating by parts the second member leads to
\begin{equation}
\int_0^R \der r\,r^2\,\left[\rho(r)-\rho'(r)\right]\,J(r)=0\,,
\label{A4}
\end{equation}
where we have introduced the definition
\begin{equation}
J(r)\equiv \int_0^r\,\der \xi\,\frac{M(\xi)+M'(\xi)}{\xi^2}\,.
\label{A5}
\end{equation}
Equation (\ref{A4}) can be recast in the form
\begin{eqnarray}
\int_0^{r\eq} \der r\,r^2\,\left[\rho(r)-\rho'(r)\right]\,J(r)=
\nonumber \\
\int_{r\eq}^R \der r\,r^2\,\left[\rho'(r)-\rho(r)\right]\,J(r)\,,
{~~~~}\label{A6}
\end{eqnarray}
with the integrands on both sides of equation (\ref{A6}) being two
strictly positive (or negative) functions in the respective open
intervals $(0,r\eq)$ and $(r\eq,R)$. Since $J(r)$ is a strictly
outwards increasing positive function, $J(r\eq)$ is, at the same time,
an upper bound for $r\in [0,r\eq]$ and a lower bound for $r\in
[r\eq,R]$. Thus, we have the following inequalities
\begin{eqnarray}
\int_0^{r\eq} \der r\,r^2\,[\rho(r)-\rho'(r)]\,J(r) < \nonumber \\
\int_0^{r\eq} \der r\,r^2\,[\rho(r)-\rho'(r)]\,J(r\eq) {~~~~}
\label{AB2ba}
\end{eqnarray}
\begin{eqnarray}
\int_{r\eq}^R \der r\,r^2\,[\rho'(r)-\rho(r)]\,J(r) > \nonumber \\
\int_{r\eq}^R \der r\,r^2\,[\rho'(r)-\rho(r)]\,J(r\eq)\,,
\label{AB2b}
\end{eqnarray}
which, given equation (\ref{A6}), lead to
\begin{eqnarray}
J(r\eq) \int_0^{r\eq} \der r\,r^2\,[\rho(r)-\rho'(r)] > \nonumber \\
J(r\eq) \int_{r\eq}^R \der r\,r^2\,[\rho'(r)-\rho(r)]  {~~~~}
\label{AB3}
\end{eqnarray}
or, equivalently, to
\begin{equation}
\int_0^{R} \der r\,r^2\,\rho(r)>
\int_{0}^R \der r\,r^2\,\rho'(r)\,.
\label{ine}
\end{equation}
As inequality (\ref{ine}) contradicts the fact that the two density
profiles must yield the same total mass $M$, the initial postulate
that they coincide at one single radius $r\eq$ can be rejected.

Two profiles of the NFW-like form, equation (1), yielding the same
mass $M$, are either identical or intersect at one single intermediate
radius $r\eq$ in the interval ($0,R$). Indeed, if they are not
identical, they have different values of $\rs$ and of the
characteristic density $\roc$, implying that they have different
slopes at all radii. This ensures that the function $\rho(r)-\rho'(r)$
has no extreme in the interval $(0,R)$ and, consequently, that it only
vanishes once. Thus, according to the previous proof, we conclude
that, if at a given epoch two (spherical, non-rotating) relaxed haloes
have identical values of $R$, $M$, $E$ and $P$, their density profiles
of the NFW universal form must necessarily coincide, too.

\section{The $\mss$ vs. $\rss$ Scaled Correlation}
\label{correlation}

\begin{figure}
\centerline{\includegraphics*[width=9cm]{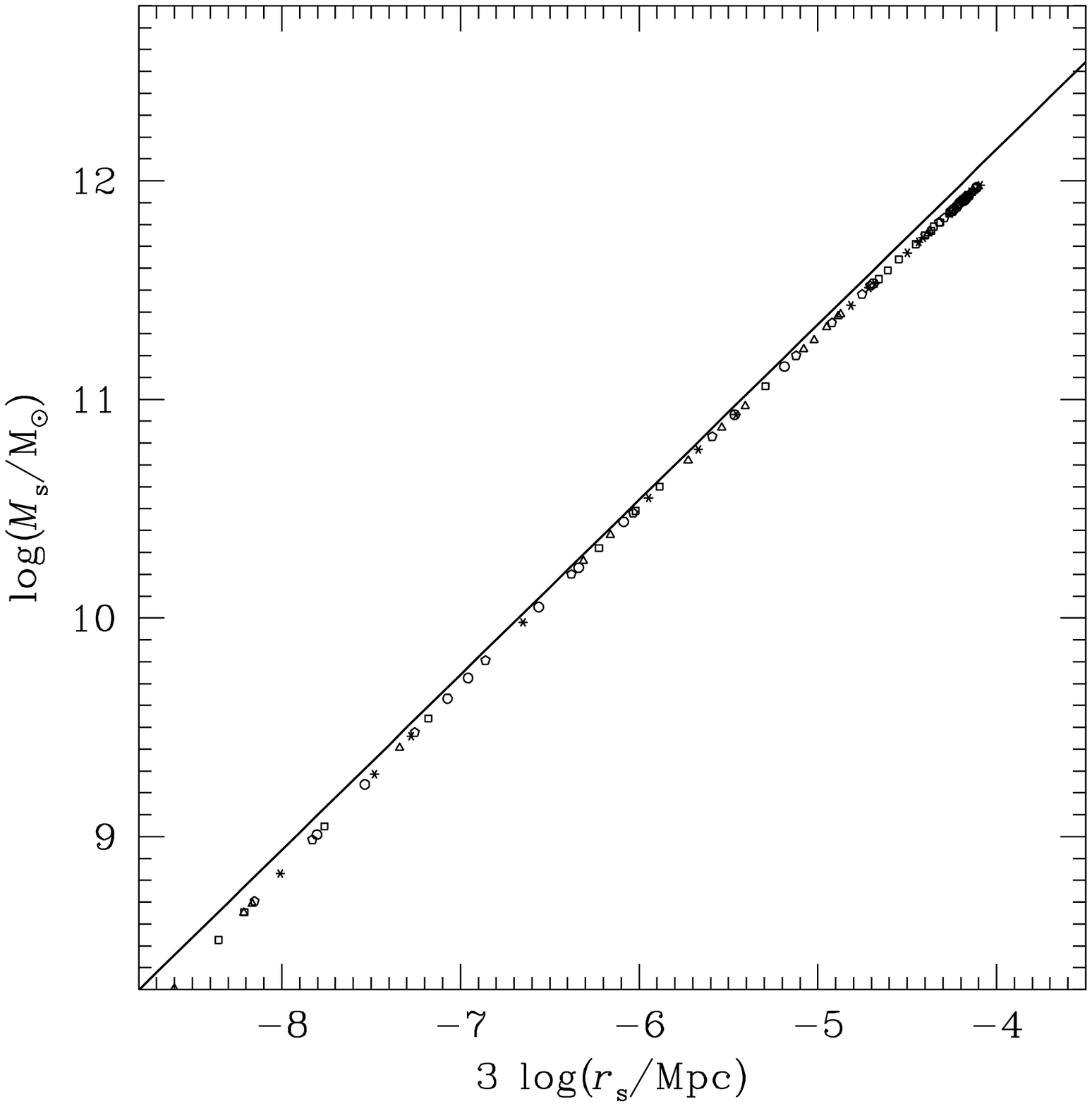}}
\caption{Typical $\rs - M\s$ relation obtained from the same haloes
as in Figures \ref{5} and \ref{6} (same symbols as in
Figure \ref{7}) without adding any scatter in $c$ and prior to the
scaling of $\rs$ and $M\s$ and the subsequent splitting in two
relations on both sides of the reference halo. To better appreciate
the small bending of this relation we show its log-log linear fit
(full line) shifted vertically by 0.07.}
\label{8}

\bigskip
\end{figure}

Let us neglect, for simplicity, the deviates of $E$, $\dot M$, and
$\dot E$ of individual haloes from their typical values dependent on
$M$ and $t$. As explained in \S~\ref{model}, the $\rs - M\s$
relation can be inferred from the mass-concentration relation at any
given redshift and equation (\ref{msm}) arising from the universal
density profile of haloes. The mass-concentration relation at $z=0$
given by \citet{Bull01} for the same cosmology as in WBPKD and ZMJB
leads to the nearly linear $\rs - M\s$ relation in log-log units
shown in Figure \ref{8}. Should this relation be strictly linear, the
$\rs$ and $M\s$ values of haloes would satisfy the relation
\begin{equation}
\log M\s = \alpha_0 + 3 \alpha_1 \log \rs\,.
\label{AA3}
\end{equation}
Since the values $\rso$ and $M\so$ of any {\it arbitrary\/} halo
eventually used to scale the $\rs$ and $M\s$ values of the different
objects would also satisfy the relation (\ref{AA3}), we are led to the
fact that haloes also show a perfect linear relation (with no scatter)
of the form
\begin{equation}
\log (M\s/M\so) = 3 \alpha_1 \log (\rs/\rso)\,.
\label{AA4}
\end{equation}
However, the real $\rs - M\s$ relation shown in Figure \ref{8}
shows a small bending so that haloes rather satisfy the relation
\begin{equation}
\log M\s = \alpha_0 + 3\alpha_1 \log \rs + 3 \alpha_2 (\log \rs)^2\,,
\label{AA45}
\end{equation}
being $\alpha_1>0$, $\alpha_2<0$ and
$|\alpha_2\,\log(\rsmax/\rsmin)|\ll \alpha_1$, where $\rsmax$ and
$\rsmin$ are the upper and lower bounds of $\rs$ in the sample. Given
that the values $\rso$ and $M\so$ of any reference halo used to scale
$\rs$ and $M\s$ also satisfy the relation (\ref{AA45}), we are led to
the scaled relation
\begin{eqnarray}
\log (M\s/M\so) = 3 \alpha_1 \log (\rs/\rso) \hspace{16truemm} \nonumber \\
+ \,\, 3\alpha_2\,[(\log \rs)^2-(\log \rso)^2]\,,
\end{eqnarray}
which can be rewritten in the form
\begin{eqnarray}
\log (M\s/M\so) = 3 \alpha_1 \left(1+ 2\,
\frac{\alpha_2}{\alpha_1} \log \rso\right) \log (\rs/\rso)  \nonumber \\
+ \, 3 \alpha_2 \,[\log (\rs/\rso)]^2\,.\hspace{22truemm} 
\label{AA5}
\end{eqnarray}
Since $|\alpha_2\,\log(\rs/\rso)|\ll \alpha_1$, the relation
(\ref{AA5}) is well approximated by the linear relation
\begin{eqnarray}
\log (M\s/M\so) = 3 \alpha_1\,\log (\rs/\rso)  {~~~~~~~~~}\nonumber \\  
\times \left\{1+ \frac{\alpha_2}{\alpha_1}
\left[2\log \rso + \log \left(\frac{\rsm}{\rso}\right)\right] \right\}
\,, 
\label{AA6}
\end{eqnarray}
where $\rsm$ is some intermediate value between $\rso$ and $\rsmax$ or
$\rsmin$ depending on whether $\rs$ is larger or smaller than $\rso$,
respectively. Thus, the slope depends, in this case, on the specific
value of $\rso$ used to scale $\rs$. Consequently, by taking $\rso$
equal to the scale radius of some halo along the typical or individual
MAT of each object, the slope will slightly change, yielding
some scatter in the scaled relation. Note however that this scatter
should be very small as $|\alpha_2\,\log(\rsm/\rso)/\alpha_1|$ is
always much smaller than unity.  On the other hand, since $\rsm/\rso$
is then respectively either smaller or larger than unity, given the signs of
coefficients $\alpha_1$ and $\alpha_2$, the slope of the scaled
correlation will be systematically larger, though by a small amount,
for $\rs\le \rso$ than in the case $\rs>\rso$.

Finally, we do not expect any noticeable difference between the $z\fa$
and $z\tp$ variants of the scaled correlations. As mentioned in \S
\ref{sim}, the values of $z\fa$ along individual MATs are of the same
order as $z\tp$, so their respective values of $\rso$ should be
similar. Certainly, the redshifts $z\fa$ correspond to objects along
typical MATs, but individual MATs are not very different from the
latter. Thus, both variants of the scaled correlations should show
essentially the same slope and scatter.

\end{document}